\documentclass[runningheads]{llncs}
\usepackage{makecell}
\usepackage[T1]{fontenc}
\usepackage{graphicx}
\usepackage{subcaption}

\usepackage{booktabs}
\usepackage{url}
\usepackage{multirow}
\begin{document}
\title{StallGrid: Measuring Internet-exposed Engagement in Protocol-Native OT Tarpits}
\titlerunning{Measuring Internet-exposed Engagement in Protocol-Native OT Tarpits}
\author{Arthur Cordeiro \and
Casper Andersen \and
Emmanouil Vasilomanolakis
}
%
% \authorrunning{C. Andersen et al.}
\institute{Technical University of Denmark, Lyngby, Denmark\\
\email{\{arur,emmva\}@dtu.dk}\\ \email{casperja@outlook.com}}
\maketitle              % typeset the header of the contribution
\begin{abstract}
Operational technology systems face exposure to scanning and protocol-specific attack tools, where compromise risks disrupting physical processes rather than just data. Traditional OT defenses rely on blocking and filtering under strict patch constraints, while tarpitting delays scanners through sustained protocol-level interaction. Modbus TCP and IEC-104 carry no native authentication or integrity protection. Application-layer tarpitting, which stalls scanners inside a legitimate protocol exchange, has been studied for IT and IoT protocols, but not for OT/ICS, whose session semantics (state machines, exception codes) create stalling opportunities IT/IoT lack, and whose patch-constrained environments need exactly this kind of alternative defense. We present StallGrid, to our knowledge the first application-layer tarpits for OT/ICS, stalling scanners via Modbus Exception Codes \texttt{0x05}/\texttt{0x06} and prolonged residence in IEC-104's connected state machine. Five variants (three Modbus TCP, two IEC-104) ran simultaneously for 24 days online, logging 6,709 sessions, 2,039 cumulative per-tarpit unique IPs, and over 2,000 hours of accumulated connection engagement. GreyNoise enrichment attributes 87--97\% of stall time to malicious-tagged IPs, just 22--30\% of connecting addresses; protocol-level behavior further correlates with malicious classification, a fingerprinting signal beyond raw stall time. Shorter induced delay (1.5s) yielded more total engagement than longer delay (3s), observed across both protocols. These results position protocol-native tarpitting as a practical, low-cost complementary defense for OT/ICS environments where patching remains infeasible.
\end{abstract}

\section{Introduction}\label{sec:intro}

Attacks against operational technology (OT) and industrial control systems (ICS) represent a latent risk. FrostyGoop manipulated heating controllers over Modbus~\cite{dragos2024frostygoop_intel}; Industroyer~\cite{eset2017industroyer} and CosmicEnergy~\cite{mandiant2023cosmicenergy} implemented IEC 60870-5-104 (IEC-104) command sequences to operate breakers in electrical power grids; Triton targeted a safety instrumented system~\cite{mandiant2017triton}. Several of these campaigns are tied to sabotage OT systems, and the capability they demonstrate is specific rather than generic: purpose-built code speaking industrial protocols to disrupt physical processes. % Modbus and IEC-104 recur across this set, making defensive measures for these two protocols a concrete priority for infrastructure availability and safety.

The early tarpit studies were narrow responses to specific threats: LaBrea slowed the CodeRed worm with zero-window TCP replies~\cite{liston2001labrea}, the Linux Netfilter TARPIT target generalized the trick at the transport layer~\cite{morris2006netfiltertarpit}, distributed SMTP tarpitting delayed protocol acknowledgements to slow spam~\cite{hunter2003distributedtarpitting}, and Malpity automatically exploited vulnerabilities in specific malware binaries to trap infected hosts~\cite{walla2019malpity}. Transport-layer tarpits are deterministic and protocol-agnostic, but that same regularity makes them easier fingerprintable: Degreaser identifies them from TCP-level artifacts at internet scale~\cite{uncoveringNetTar}, and the countermeasures proposed in response~\cite{shing2016improved} introduce detectable artifacts of their own, sustaining an arms race with no stable advantage for the defender. Even recent transport-layer tarpits built against IoT worm propagation face the same structural exposure~\cite{griffioen2023pitoftar}. This motivated a shift toward application-layer tarpits, which stall inside a legitimate protocol exchange and therefore require the scanner to possess protocol knowledge and maintain session state to disengage. Endlessh~\cite{WellonsEndlessh} demonstrated the idea for SSH banner exchange, and Franco et al. survey the broader IoT, IIoT, and cyber-physical-systems deception literature~\cite{franco2021honeypotsurvey}. EventHorizon~\cite{safargalieva2026eventhorizon} recently tested the approach empirically: an eight-week global deployment of multiprotocol application-layer tarpits covering Telnet, MQTT, UPnP, and CoAP. Application-layer tarpitting is now an empirically studied defensive technique. The evidence base, however, lies entirely in the generic IoT space.

OT/ICS differs from IoT along the dimensions that govern both tarpit design and defensive value. Modbus and IEC-104 carry no native authentication or integrity protection~\cite{modbus_protocol_2012,ds_en_60870_5_104}, patching is constrained by availability requirements and equipment lifecycles measured in decades~\cite{nist80082r3,schachinger2026purity}, and the consequence of compromise is physical process disruption with safety implications rather than data loss or botnet recruitment~\cite{attackvectorsSCADA,ALCARAZ201553}. This is not merely theoretical: Yaben et al. find, via internet-wide measurement, that a substantial share of internet-facing OT and IoT devices are neglected, running obsolete software with no realistic maintenance path~\cite{yaben2024neglectedobsolete}. Conventional countermeasures compound the problem: botnet takedowns and blacklisting are short-lived once infrastructure relocates~\cite{georgoulias2023botnetbusiness}. No prior work designs, deploys, or empirically evaluates application-layer tarpits for OT protocols. % IoT results do not transfer by assumption. OT protocols expose session semantics that generic IoT protocols lack: Modbus Exception Codes signal recoverable device conditions that a client is expected to tolerate~\cite{modbus_protocol_2012}, and IEC-104 maintains an explicit connected state machine~\cite{ds_en_60870_5_104}. Both create protocol-native stalling opportunities with no IoT analogue. The open question is therefore whether tarpitting is a practical defensive layer for industrial protocols specifically, because the OT environment admits few alternatives: operators who cannot patch quickly or deploy heavyweight security stacks need low-cost complementary options.

This paper answers that question empirically. We design, implement, and deploy application-layer tarpits for Modbus and IEC-104 on the open internet, and evaluate them against live scanning traffic enriched with independent threat intelligence. In summary, the contributions of our paper are as follows:

\begin{itemize}
    \item \textbf{OT-specific application-layer tarpits.} We design and implement tarpits for Modbus and IEC-104 that stall using protocol-native semantics: Modbus Exception Codes \texttt{0x05} and \texttt{0x06}, and prolonged residence in the IEC-104 connected state without promising data delivery. To the best of our knowledge these are the first application-layer tarpits targeting OT/ICS protocols.

    \item \textbf{Live empirical deployment.} We deploy five tarpit variants simultaneously for 24 days: three Modbus configurations (1.5\,s delay with \texttt{0x05} and \texttt{0x06} replies, 3\,s delay with the same replies, and 1.5\,s delay with \texttt{0x06} only) and two IEC-104 configurations (1.5\,s and 3\,s delay), each on a dedicated public IP. Together they attract several thousand sessions and accumulate over 2{,}000 cumulative hours (approximately 83 days) of scanner engagement time.

    \item \textbf{Threat-intelligence-corroborated effectiveness.} Enriching every connecting address with GreyNoise tags shows that roughly 87--97\% of stall time, depending on the variant, is attributable to independently classified malicious-tagged IPs. Protocol-level behavior, specifically Modbus function-code distribution and IEC-104 protocol-compliance violations, correlates with that classification, yielding fingerprintable defensive signal beyond raw time consumption.

    \item \textbf{Delay/engagement trade-off.} We characterize how induced reply delay shapes scanner behavior. Shorter delays (1.5\,s) accumulate more total stall time and more connections per address than longer delays (3\,s) in both protocols -- but not more unique scanners: IEC-104's 3\,s tarpit in fact drew more unique IPs (605) than its 1.5\,s tarpit (338). Longer delays instead show lower aggregate time per session and are measurably harder for common scanning tools to register as a live protocol responder. This gives an actionable parameterization trade-off for future OT tarpit deployments.
    
    \item \textbf{Experiment Artifacts.} Both tarpits are implemented as standalone, epoll-based servers and released as reference artifacts with the anonymized collected data (Appendix~\ref{sec:open-science}).\footnote{\url{https://anonymous.4open.science/r/StallGrid-A765/}}
\end{itemize}

The remainder of the paper proceeds as follows.  Section~\ref{sec:related-work} reviews prior work; Section~\ref{sec:design} and Section~\ref{sec:eval-overview} present our design and deployment; Section~\ref{sec:eval-results} reports results; Section~\ref{sec:discussion} discusses findings, Section~\ref{sec:limitations} covers limitations and mitigations, and Section~\ref{sec:conclusion} concludes and suggests future directions.

\section{Related Work}
\label{sec:related-work}

Transport-layer tarpits established stalling as a defensive primitive. LaBrea~\cite{liston2001labrea}, the Netfilter TARPIT target~\cite{morris2006netfiltertarpit}, and distributed SMTP tarpitting~\cite{hunter2003distributedtarpitting} hold a connection open beneath the application protocol, regardless of what the connecting party is trying to speak. That uniformity is also their weakness. Degreaser separates them from real hosts at internet scale~\cite{uncoveringNetTar}, the countermeasures proposed in response carry artifacts of their own~\cite{shing2016improved}, and recent transport-layer deployments against IoT worm propagation inherit the same exposure~\cite{griffioen2023pitoftar}. Application-layer tarpits answer this by stalling inside a protocol exchange the scanner must understand to disengage, which forces it to hold session state for as long as it stays. Malpity~\cite{walla2019malpity} and Endlessh~\cite{WellonsEndlessh} demonstrated the mechanism, and EventHorizon~\cite{safargalieva2026eventhorizon} delivered the first internet-scale empirical validation of it, spanning Telnet, MQTT, UPnP, and CoAP. Every application-layer tarpit tested so far therefore speaks either an IT protocol such as SMTP or SSH, or an IoT one. None has been built or tested for an industrial protocol, and that is the space this paper occupies.

OT is hardware-limited and hard to maintain: legacy control systems resist frequent patching and demand real-time availability guarantees stricter than those of conventional IT~\cite{cardenas2008research}. Security has to adapt to that constraint, not the other way around. The deception work that does target OT reflects this constraint rather than escaping it. Ramachandruni and Poornachandran deploy a Modbus honeypot against SCADA attack vectors~\cite{attackvectorsSCADA}, and Grigoriou et al. build one for IEC 60870-5-104~\cite{grigoriou2022protecting}. Both accept and log traffic passively instead of stalling it. They observe an attacker. They do not impose cost on one. Passive imitation on its own is not sufficient either. Mladenov et al. scan the internet for exposed industrial honeypots and fingerprint them at rates reaching 92\% in some protocol categories~\cite{glittersGold}. A system an attacker can recognize as a decoy has already lost whatever advantage it was deployed for. Yaben et al. measure how OT networks sit exposed on the public internet~\cite{yaben5974783measuring}, and the population they find is already large. This environment needs a defense that operates within its hardware and resource limits rather than against them.

We introduce the first protocol-aware, application-layer tarpit framework targeting Modbus and IEC-104. For each protocol, we derive stalling behaviour from interaction primitives already defined by the standards~\cite{modbus_protocol_2012,ds_en_60870_5_104}, namely Modbus Exception Codes 0x05 and 0x06 as well as the IEC-104 connected state. We implement dedicated tarpit behaviours from these primitives that go beyond generic transport-level throttling. In contrast to prior OT deception work, which passively observes and logs attacker activity, our approach is tailored to concrete protocol semantics and imposes engagement cost directly. Importantly, by covering both Modbus's transactional request-response exchange and IEC-104's persistent connected session, we also address two distinct session models and evaluate scanner engagement with each under real-world scanner interaction.

No prior system combines OT-protocol specificity, active stalling that imposes engagement cost, live internet deployment, and independent threat-intelligence corroboration (Table~\ref{tab:related-comparison}); this paper closes that combined gap.

\begin{table}[ht]
    \centering
    \small
    \setlength{\tabcolsep}{4pt}
    \caption{Positioning against the closest related deception systems.}
    \label{tab:related-comparison}
    \begin{tabular}{@{}lcccc@{}}
        \toprule
        \textbf{System} & \makecell{\textbf{OT}\\\textbf{protocol}} & \makecell{\textbf{Imposes}\\\textbf{stall cost}} & \makecell{\textbf{Live}\\\textbf{deployment}} & \makecell{\textbf{Threat-intel}\\\textbf{corroborated}} \\
        \midrule
        Modbus honeypot~\cite{attackvectorsSCADA} & Yes & -- & Yes & -- \\
        IEC-104 honeypot~\cite{grigoriou2022protecting} & Yes & -- & Yes & -- \\
        EventHorizon~\cite{safargalieva2026eventhorizon} & -- & Yes & Yes & Yes \\
        \textbf{This work} & Yes & Yes & Yes & Yes \\
        \bottomrule
    \end{tabular}
\end{table}
\section{Design}\label{sec:design}

The system comprises two independent tarpits, one per protocol, each emulating the passive server-side role: the Modbus server and the IEC-104 controlled station, accepting only client-initiated connections and never originating traffic. Each runs as a dedicated process implemented in C for deterministic response timing under many concurrent connections. Both share one architectural principle: the stall is produced entirely from protocol-native semantics that keep every exchange specification-compliant, rather than from transport-layer manipulation such as TCP zero-window signaling~\cite{rowe2026designing}.

\subsection{Modbus Tarpit}
\begin{figure}[!htb]
    \centering
    \begin{subfigure}[b]{0.49\textwidth}
        \centering
        \includegraphics[width=\textwidth]{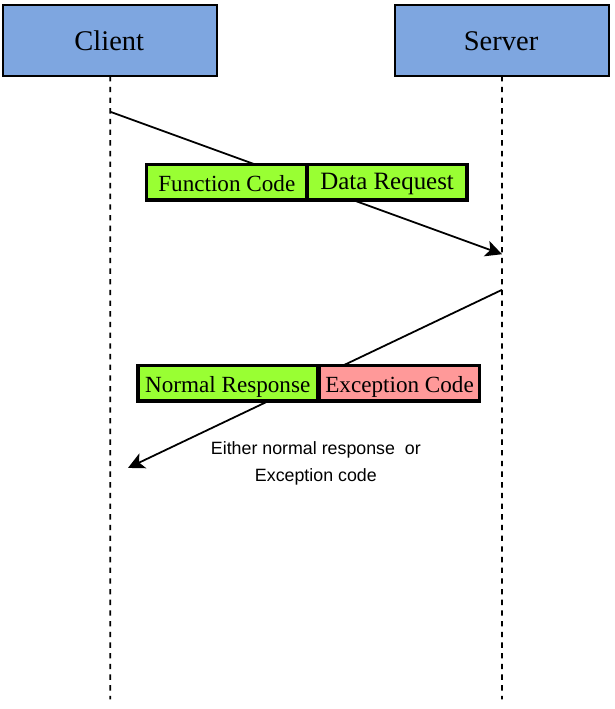}
        \caption{Common chronologic schema of a Modbus client-server model.}
        \label{fig:modbus-exception}
    \end{subfigure}
    \hfill
    \begin{subfigure}[b]{0.5\textwidth}
        \centering
        \includegraphics[width=\textwidth]{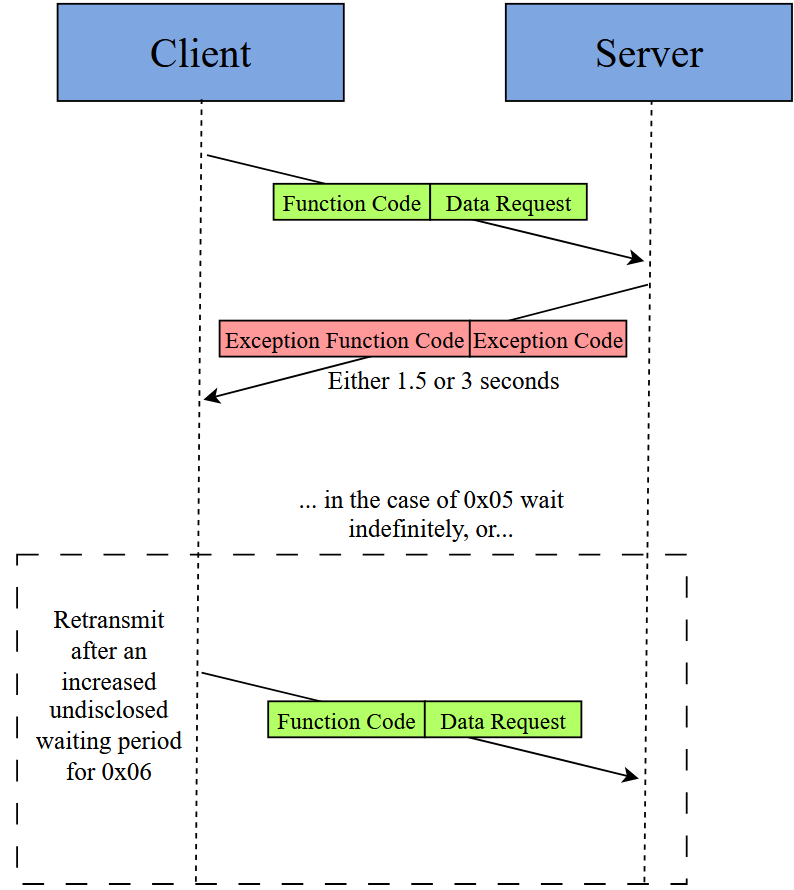}
        \caption{Our implemented Exception-code strategy stalling with \texttt{0x05}/\texttt{0x06} codes.}
        \label{fig:modbus-fc43}
    \end{subfigure}
    \caption{Modbus request-response behavior and the implemented exception-code stalling strategy under bounded 1.5-/3-second response delays.}
    \label{fig:modbus-strategies}
\end{figure}

Modbus is strictly request-response: the server may transmit only in reply to a client request, never unsolicited~\cite{modbus_protocol_2012}. The tarpit exploits this by answering every request not with the requested register, coil, or diagnostic data, but with a Modbus exception response. The specification defines two exception codes whose semantics instruct the client to keep waiting rather than abandon the exchange: \texttt{0x05} and \texttt{0x06} (Fig.~\ref{fig:modbus-exception})~\cite{modbus_protocol_2012}. Before issuing either, the tarpit inserts a configurable delay sized relative to the approximately two-second default per-request timeout used by common Modbus scanning tools (e.g., Nmap, Metasploit), maximizing socket lifetime without crossing the abandonment threshold. The two codes compose, singly or in sequence: an initial \texttt{0x05} reply followed by \texttt{0x06} on all subsequent requests is one example. A Read Device Identification alternative (Function Code 43) was also considered, using the \texttt{More Follows} field to invite indefinite chunked retrieval~\cite{modbus_protocol_2012}, but was not implemented in the current system (Fig.~\ref{fig:modbus-fc43}).

\paragraph{Acknowledge.}
Code \texttt{0x05} is defined to mean that the server has accepted the request and is processing it, but that a long duration will be required. This response exists to prevent a client timeout~\cite{modbus_protocol_2012}. Issuing it licenses the client to keep the connection open indefinitely, awaiting a promised reply the tarpit never sends. The specification's own wording removes the client's incentive to disconnect.

\paragraph{Server Device Busy.}
Code \texttt{0x06} is defined to mean that the server is engaged in a long-duration program command and that the client should retransmit later, once free~\cite{modbus_protocol_2012}. The tarpit answers every retransmission with the same code and keeps no additional state. The client's own spec-compliant retry behavior sustains the exchange.

\subsection{IEC-104 Tarpit}

\begin{figure}[!htb]
    \centering
    \begin{subfigure}[b]{0.51\textwidth}
        \centering
        \includegraphics[width=\textwidth]{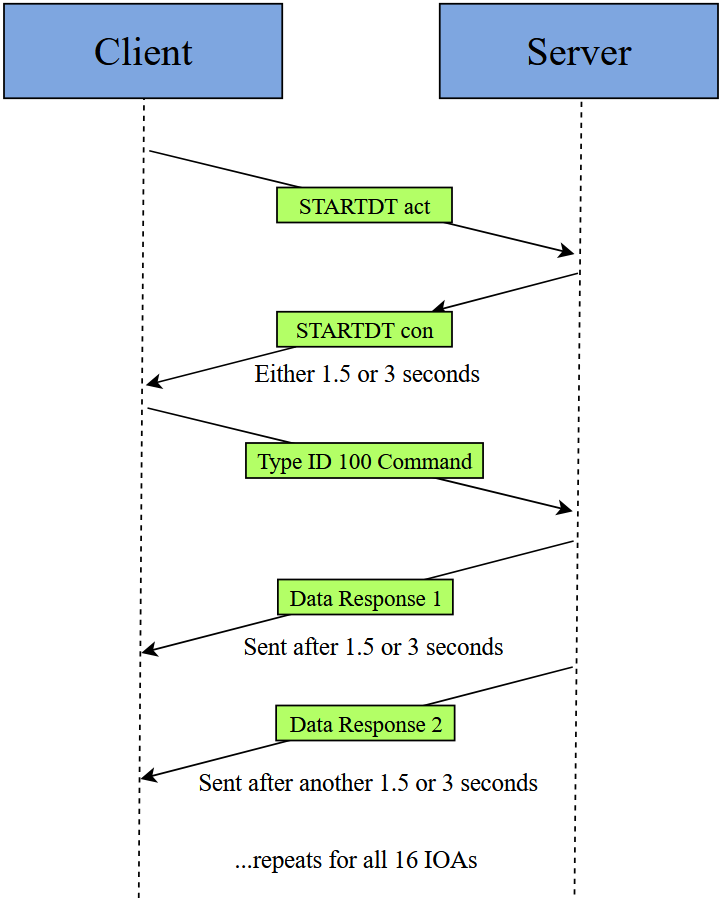}
        \caption{General Interrogation reply delayed by induced latency.}
        \label{fig:iec-latency}
    \end{subfigure}
    \hfill
    \begin{subfigure}[b]{0.48\textwidth}
        \centering
        \includegraphics[width=\textwidth]{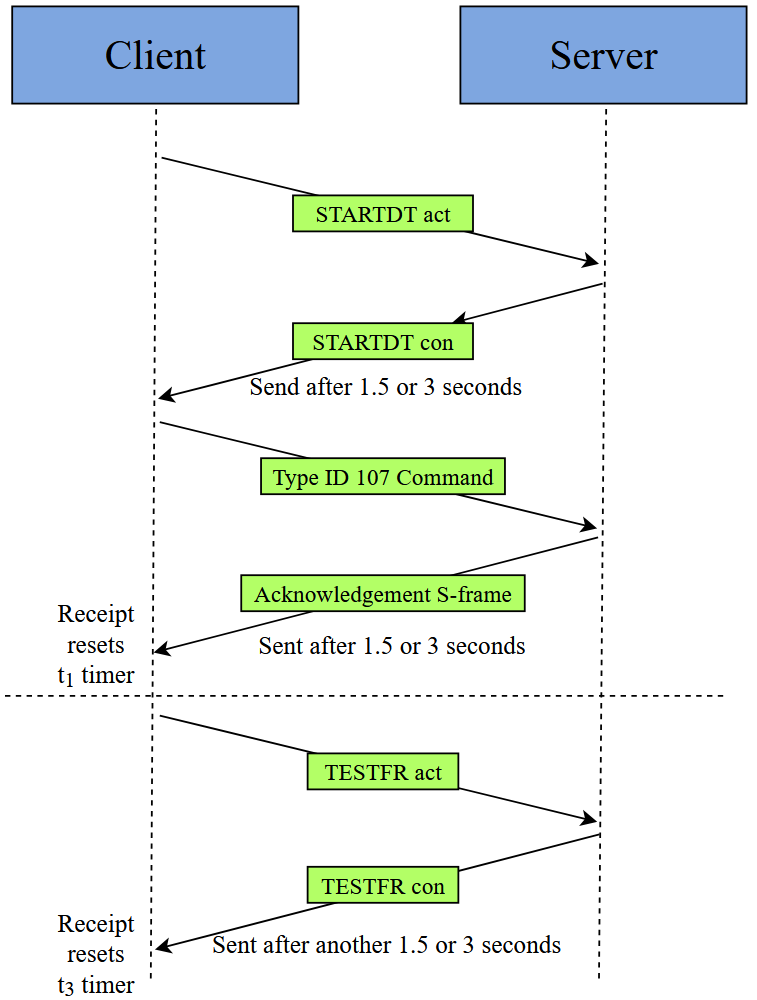}
        \caption{Supervisory acknowledgment resetting $t_1$ and $t_3$.}
        \label{fig:iec-timeout}
    \end{subfigure}
    \caption{IEC-104 stalling strategies using delayed General Interrogation responses and supervisory acknowledgments.}
    \label{fig:iec-strategies}
\end{figure}

IEC-104 is a stateful, timer-governed protocol, not a fixed request-response exchange. A connection becomes an active data-transfer session only after a \texttt{STARTDT} activation/confirmation exchange. Once active, correctness is enforced by two timers: $t_1$, the maximum wait for acknowledgment of a sent frame before the connection is deemed failed, and $t_3$, the maximum idle time before a link-test frame must confirm liveness~\cite{ds_en_60870_5_104}. The tarpit completes the \texttt{STARTDT} handshake as the controlled station. It then answers the client's General Interrogation command (Type ID 100, \texttt{C\_IC\_NA\_1}) with a synthetic sequence of I-frames describing information objects of a simulated substation (Fig.~\ref{fig:iec-latency}), following a standard 110~kV topology used in related IEC-104 intrusion-detection research~\cite{Egger2020}. For every subsequent command, the tarpit withholds the corresponding data and returns only a supervisory (S-format) APDU, which the standard defines as a receipt acknowledgment carrying no information object.

\paragraph{Supervisory Acknowledgment.}
Sending the S-format APDU resets the client's $t_1$ timer without conveying any information object. The link stays formally valid regardless. To survive the $t_3$ interval, the tarpit answers \texttt{TESTFR} activation frames with a \texttt{TESTFR} confirmation, without exchanging application data. Both timers that would otherwise terminate the connection, $t_1$ and $t_3$, are continuously neutralized (Fig.~\ref{fig:iec-timeout}). The General Interrogation reply remains the only substantive data ever transmitted, and the connection's state machine stays valid at every layer the client checks even as it makes no further application-layer progress~\cite{ds_en_60870_5_104}

\section{Implementation Overview}
\label{sec:eval-overview}

\subsection{Tarpit Implementations}
\label{sec:eval-tarpits}

The induced delays were fixed at $1.5$\,s and $3$\,s, bracketing the $\approx2$\,s per-request timeout assumed by common Modbus scanning tools ~\cite{nmapModbusDiscover}\cite{metasploit2026modbusclient}. IEC-104 has no such fixed client default; its connections instead abandon on the $t_1$/$t_3$ timers~\cite{ds_en_60870_5_104}. The same $1.5$\,s/$3$\,s pair was reused there regardless, for cross-protocol comparability. Deployment IP availability limited the study to five tarpits: three Modbus, two IEC-104 (Table~\ref{tab:tarpitsNumbers}). Tarpits 1 and 2 share an identical \texttt{0x05}$\to$\texttt{0x06} reply strategy and vary only the configured delay magnitude. Tarpit 5 repeats Tarpit 1's $1.5$\,s delay but replies with \texttt{0x06} only, varying the initial acknowledgment code instead. Tarpits 3 and 4 apply the same $1.5$\,s/$3$\,s contrast to IEC-104. Each tarpit runs on its own public IP with no rotation or replication of a given delay configuration across IPs (Section~\ref{sec:eval-setup}), so these pairings support observed associations between delay and scanner behavior, not an isolated causal effect of delay magnitude alone.

\begin{table}[ht]
    \centering
    \begin{tabular}{@{}cllc@{}}
        \toprule
        \textbf{Tarpit ID \#} & \textbf{Protocol} & \textbf{Strategy} & \textbf{Response Delay} \\
        \midrule
        1 & Modbus  & Exception Code (\texttt{0x05}$\to$\texttt{0x06}) & 1.5\,s \\
        2 & Modbus  & Exception Code (\texttt{0x05}$\to$\texttt{0x06}) & 3\,s \\
        3 & IEC-104 & Supervisory Acknowledgment         & 1.5\,s \\
        4 & IEC-104 & Supervisory Acknowledgment         & 3\,s \\
        5 & Modbus  & Exception Code (\texttt{0x06} only) & 1.5\,s \\
        \bottomrule
    \end{tabular}
    \caption{Deployed tarpit implementations: protocol, stalling strategy, and induced delay.}
    \label{tab:tarpitsNumbers}
\end{table}

\subsection{Metrics}
\label{sec:eval-metrics}

Three categories quantify tarpit behavior. \textit{Effectiveness} covers connection engagement and intelligence gathered. We report \textit{stall time}, the connection-open duration summed across sessions, as the primary engagement measure, with concurrent-connection count as a secondary signal; no non-tarpitted control condition was deployed, so stall time is reported as an absolute accumulated-connection-time quantity rather than as a reduction relative to a baseline (Section~\ref{sec:eval-results} treats cross-configuration comparisons accordingly). An open connection is counted as one occupied socket, not necessarily one continuously occupied scanner-side CPU thread; the frequency of client requests supplements this as a coarser activity signal. Network-overhead benchmarking, relevant to higher-frequency IoT threats, falls outside OT's low-frequency reconnaissance scope and is not measured. Intelligence gathered comes from requests-per-client and concurrent-client counts as proxies for probing intent, and from protocol/client state-machine integrity as a behavioral signal.

\textit{Practicality}, measures whether a tarpit survives the protocol-level handshake check, a valid MBAP header or a \texttt{STARTDT} confirmation, that scanners use to discard non-devices. It is also read from the ratio of silent to requesting connections and their respective durations, which indicates selective engagement of protocol-aware scanners, and from the preference for application-layer stalling over transport-layer manipulation such as TCP zero-window, safer for legacy OT networks.

\textit{Defensive role}, covers the capture of scanner commands for behavioral fingerprinting and the detection of protocol violations, out-of-order packets, invalid sequence numbers, malformed function codes, as a high-fidelity non-compliance signal. It also covers the fusion of captured network data with external threat intelligence, chosen as an independent, third-party reputation source rather than a bespoke classifier to avoid circularity with the tarpit's own engagement metrics, toward actionable situational awareness beyond simple IP blocking. We treat its tags as corroboration rather than ground truth throughout (Section~\ref{sec:discussion}).

\subsection{Setup}
\label{sec:eval-setup}

All five tarpits were hosted on individual DigitalOcean instances in Santa Clara, California, exposing each instance directly to internet-wide scanning. Tarpit 1 went live on May 9, Tarpits 2--4 on May 14, and Tarpit 5 on May 20. All five then ran simultaneously for 24 days before being taken down on June 12. The Modbus tarpits accumulated unequal total uptime, so the subsequent analysis additionally controls for the overlapping period during which all three were live concurrently.

Before any tarpit was exposed to the public internet, an internal pre-deployment pass confirmed both that external scanning tools would fingerprint each instance as a genuine device and that a compliant protocol client could complete the expected interaction sequence. Nmap~\cite{nmap_nse_docs}, Metasploit~\cite{metasploit2026modbusclient}, and Zgrab2~\cite{durumeric2024ten}, the same tool population assumed of scanners in the wild, probed each instance during this stage. The Modbus tarpits were further exercised against the \texttt{pymodbus}~\cite{pymodbus2026} client library and the IEC-104 tarpits against \texttt{lib60870-C}~\cite{lib60870c2026} (via ctypes), covering handshake, keep-alive (\texttt{TESTFR}), supervisory acknowledgment, and general-interrogation sequences across twelve test cases. Round-trip timing, via Zgrab2 for Modbus, confirmed that the configured $1.5$\,s/$3$\,s stall held consistently beyond the handshake stage before any tarpit went live. Deployment complied with the scanning-ethics constraints detailed in Appendix~\ref{app:ethics}.

\section{Results}
\label{sec:eval-results}

\subsection{Key Findings}
\label{subsec:headlines}

This evaluation answers four questions, in the order posed: (RQ1) does stall time scale with the induced per-reply delay for Modbus (Subsection~\ref{subsec:modbus-results})? (RQ2) does the same hold for IEC-104 (Subsection~\ref{subsec:iec104-results})? (RQ3) how does engagement compare across the two protocols (Subsection~\ref{subsec:cross-protocol})? (RQ4) how much of that engagement is attributable to independently classified malicious-tagged sources (Subsection~\ref{subsec:threat-intel})? No non-tarpitted control was deployed (Section~\ref{sec:eval-metrics}), so RQ1 and RQ2 use the closest available same-protocol baseline: tarpits sharing an identical reply strategy and differing only in configured delay. Because each tarpit also runs on a distinct public IP with no rotation across configurations, these comparisons are reported as observed associations rather than isolated delay effects (Subsection~\ref{subsec:disc-retention} returns to this limitation). Table~\ref{tab:tarpits_overall} summarizes per-tarpit session counts, unique IP addresses, connection statistics, and total stall time over the full deployment period.

\begin{table}[ht]
    \centering
    \caption{Per-tarpit summary over the full deployment period.}
    \label{tab:tarpits_overall}
    \vspace{2pt}
    \small
    \setlength{\tabcolsep}{5pt}
    \begin{tabular}{lccrrcccc}
        \toprule
        \textbf{Protocol} & \textbf{ID} & \textbf{Sessions} & \textbf{IPs} & \textbf{CM} & \textbf{CP} & \textbf{Mean} & \textbf{Median} & \textbf{Total} \\
        \midrule
        \multirow{3}{*}{Modbus} & 1 & 1537 & 422 & 3.64 & 20 & 1{,}877\,s & 10\,s & 800.46\,h \\
         & 2 & 1092 & 352 & 3.09 & 28 & 1{,}411\,s & 10\,s & 427.89\,h \\
         & 5 & 976  & 322 & 3.03 & 26 & 2{,}136\,s & 10\,s & 577.83\,h \\
        \midrule
        \multirow{2}{*}{IEC-104} & 3 & 1345 & 338 & 3.96 & 12 & 294\,s & 12\,s & 107.38\,h \\
         & 4 & 1759 & 605 & 2.91 & 12 & 184\,s & 10\,s & 88.91\,h \\
        \bottomrule
    \end{tabular}
    \par\vspace{3pt}
    \noindent{\textbf{CM}: Mean Connections per IP. \quad \textbf{CP}: Concurrence Peak. \quad Total: Stall time.}
\end{table}

\begin{enumerate}
\renewcommand{\labelenumi}{\Roman{enumi})}
\setlength{\itemsep}{2pt}\setlength{\topsep}{3pt}

\item \emph{Delay-independent stall time, Modbus.}
  During the overlap period, the two 1.5-second Modbus tarpits (Tarpit 1, Tarpit 5) accumulate 563.1 and 577.83 hours of stall time respectively, 75.7\% and 80.3\% more than the 320.5 hours accumulated by the 3-second tarpit (Tarpit 2), despite the shorter per-reply delay.

\item \emph{Delay-independent stall time, IEC-104.}
  Tarpit 3 (1.5s) accumulates 107.38 hours of stall time against 88.91 hours for Tarpit 4 (3s), 20.7\% more, even though Tarpit 4 draws 30.7\% more sessions and 77.9\% more unique IPs.

\item \emph{Cross-protocol stall-time asymmetry.}
  Over the overlap period, IEC-104 draws 26.5\% more average sessions than Modbus (1212 vs.\ 958.3) yet accumulates 84.5\% less average stall time per tarpit (75.39 vs.\ 487.1 hours); pairwise comparisons range from 298\% to 650\% more stall time for Modbus tarpits over IEC-104 tarpits.

\item \emph{Threat-intelligence correlation.}
  Malicious-tagged IPs account for 87.3\%--97.4\% of total stall time across all five tarpits, despite comprising only 22--30\% of the unique addresses that connected to any of them (Table~\ref{tab:gn-class}).

\end{enumerate}

\subsection{Modbus Tarpit Performance}
\label{subsec:modbus-results}

For the Modbus tarpits, stall time does not scale with the induced per-reply delay. During the overlap period, the 1.5-second tarpits (Tarpit 1, Tarpit 5) accumulate 563.1 and 577.83 hours of stall time respectively, 75.7\% and 80.3\% more than the 320.5 hours accumulated by the 3-second tarpit (Tarpit 2), despite the shorter per-reply delay. Over the full deployment period, mean connection durations run 7.77 and 12.08 minutes longer, respectively, on the 1.5-second tarpits than on Tarpit 2, more than offsetting the shorter per-reply wait (Table~\ref{tab:tarpits_overall}). Mean connection duration (1{,}877--2{,}136\,s across the three tarpits) far exceeds the median (10\,s for all three, Table~\ref{tab:tarpits_overall}), indicating a heavy-tailed distribution in which a comparatively small number of long-duration sessions account for a disproportionate share of total stall time. Table~\ref{tab:modbus-summary} shows Tarpit 1 receiving the most requests over the full deployment (562) and Tarpit 5 the fewest (385), while Tarpit 5's stall time still exceeds Tarpit 2's (472 requests).

\begin{table}[ht]
    \centering
    \caption{Modbus tarpit's response, Exception Code overall distribution per tarpit.}
    \label{tab:modbus-summary}
    \vspace{2pt}
    \setlength{\tabcolsep}{20pt}
    \begin{tabular}{@{}cccc@{}}
        \toprule
        \textbf{Tarpit} & \textbf{Requests} & \textbf{EC05} & \textbf{EC06} \\
        \midrule
        1 & 562 & 420 & 141 \\
        2 & 472 & 336 & 131 \\
        5 & 385 & 0   & 381 \\
        \bottomrule
    \end{tabular}
    \par\vspace{3pt}
    \noindent{\textbf{EC05}: Exception Code \texttt{0x05} (Acknowledge). \quad \textbf{EC06}: Exception Code \texttt{0x06} (Server Device Busy).}
\end{table}

\begin{figure}[!htb]
\centering
\includegraphics[width=0.75\linewidth]{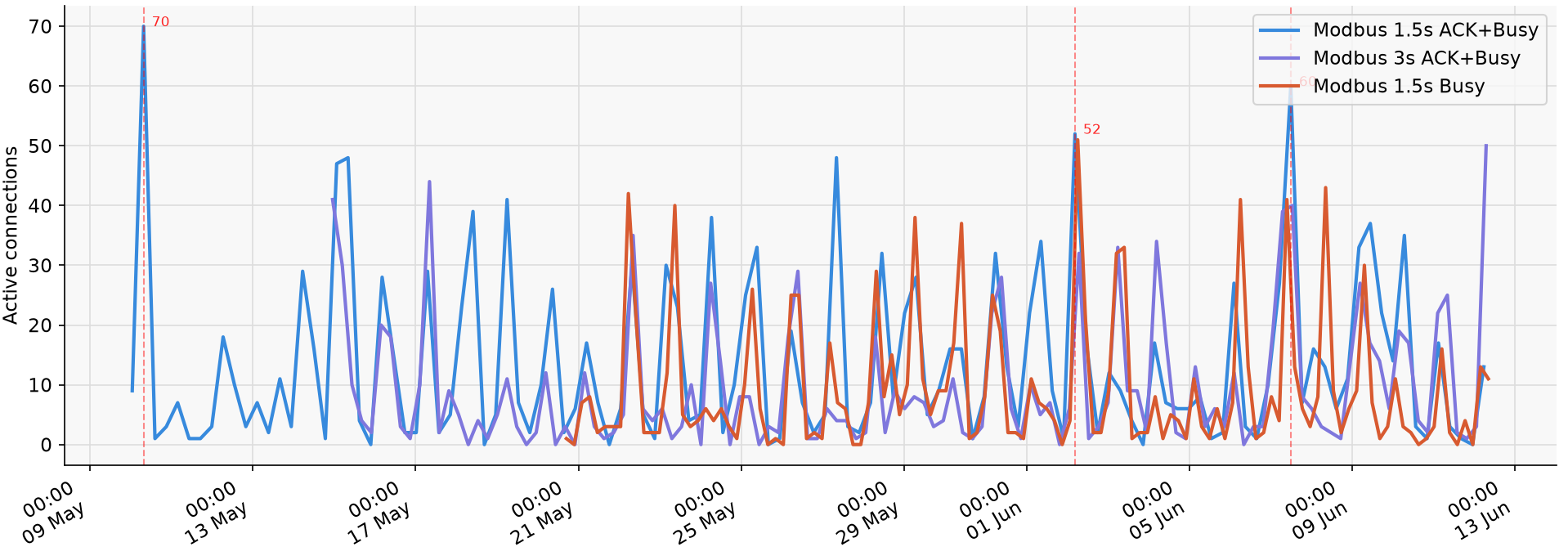}
\caption{Modbus active connections over time}
\label{fig:connections_overtime_modbus}
\end{figure}

\begin{figure}[!htb]
\centering
\includegraphics[width=0.75\linewidth]{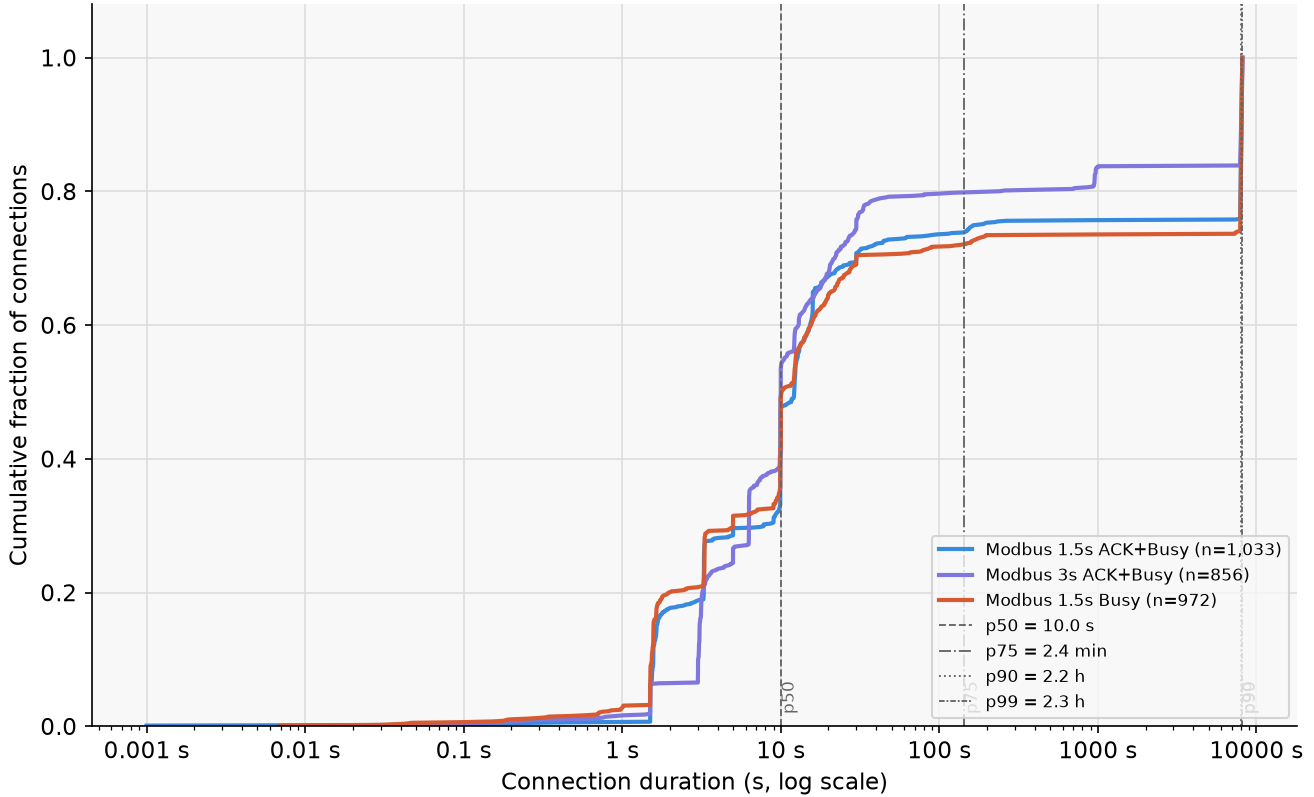}
\caption{Modbus connection-duration CDF}
\label{fig:cdf_overlap}
\end{figure}

The difference in active connections is minimal between the three Modbus tarpits, with peaks following similar, semi-synchronized patterns (Figure~\ref{fig:connections_overtime_modbus}). The distribution of connection durations, shown in Figure~\ref{fig:cdf_overlap}, has disconnect spikes around 1.5\,s for all three tarpits, with only the 3-second-delay tarpit showing a second spike near 3\,s.

Function Code 43 (Read Device Identification, MEI) accounts for the majority of Modbus requests: 72.7\% of all requests over the full deployment (1{,}031 of 1{,}419) and 72.1\% during the overlap period, against 21.4\%/22.5\% for Function Code 17 (Report Server ID) and 5.9\%/5.3\% for Function Code 03 (Read Holding Registers) (Table~\ref{tab:modbus-fc-all}).

\begin{table}[ht]
    \centering
    \caption{Modbus client's valid requests, Function Code overall distribution per tarpit.}
    \label{tab:modbus-fc-all}
    \vspace{2pt}
    \setlength{\tabcolsep}{20pt}
    \begin{tabular}{@{}cccc@{}}
        \toprule
        \textbf{Tarpit} & \textbf{FC03} & \textbf{FC17} & \textbf{FC43} \\
        \midrule
        1 & 20 (3.5\%) & 124 (22.1\%) & 418 (74.4\%) \\
        2 & 40 (8.5\%) & 98 (20.8\%)  & 334 (70.7\%) \\
        5 & 24 (6.2\%) & 82 (21.3\%)  & 279 (72.5\%) \\
        \midrule
        \textbf{Total} & \textbf{84} & \textbf{304} & \textbf{1{,}031} \\
        \bottomrule
    \end{tabular}
    \par\vspace{3pt}
    \noindent{\textbf{FC03}: Read Holding Registers. \quad \textbf{FC17}: Report Server ID. \quad \textbf{FC43}: Read Device Identification (MEI).}
\end{table}

Selective engagement is visible in the silent-versus-requesting split: only 27.3\%, 31.2\%, and 28.9\% of connections to Tarpit 1, 2, and 5 respectively ever issue a valid request, with the requesting and silent populations showing markedly different disconnect-time distributions (Figure~\ref{fig:req_vs_silent_cdf_modbus}).

\subsection{IEC-104 Tarpit Performance}
\label{subsec:iec104-results}

For the IEC-104 tarpits, stall time again does not scale with the induced per-reply delay. Tarpit 3 (1.5s) accumulates 107.38 hours of stall time against 88.91 hours for Tarpit 4 (3s), 20.7\% more, even though Tarpit 4 draws 30.7\% more sessions and 77.9\% more unique IPs. Tarpit 3 also holds a mean connection duration of 294s against Tarpit 4's 184s, 59.7\% longer (Table~\ref{tab:tarpits_overall}). Table~\ref{tab:iec104-summary} shows comparable protocol-log volumes between the two: Tarpit 3 receives 7\% more \texttt{STARTDT} commands and sends 9\% more server-initiated \texttt{TESTFR}, while Tarpit 4 receives 22\% more I-frames and 12\% more client-initiated \texttt{TESTFR}.

\begin{table}[ht]
    \centering
    \footnotesize
    \setlength{\tabcolsep}{10pt}
    \begin{tabular}{@{}ccccc@{}}
        \toprule
        \textbf{Tarpit} & \makecell{\textbf{STARTDT}\\\textbf{rcvd}} & \makecell{\textbf{I-frames}\\\textbf{rcvd}} & \makecell{\textbf{TESTFR}\\\textbf{rcvd (client)}} & \makecell{\textbf{TESTFR}\\\textbf{sent (server)}} \\
        \midrule
        3 & 139 & 45 & 108 & 445 \\
        4 & 129 & 55 & 121 & 408 \\
        \bottomrule
    \end{tabular}
    \caption{IEC-104 sessions-log summary: handshake/keep-alive counts per tarpit.}
    \label{tab:iec104-summary}
\end{table}

\begin{figure}[!htb]
\centering
\includegraphics[width=0.75\linewidth]{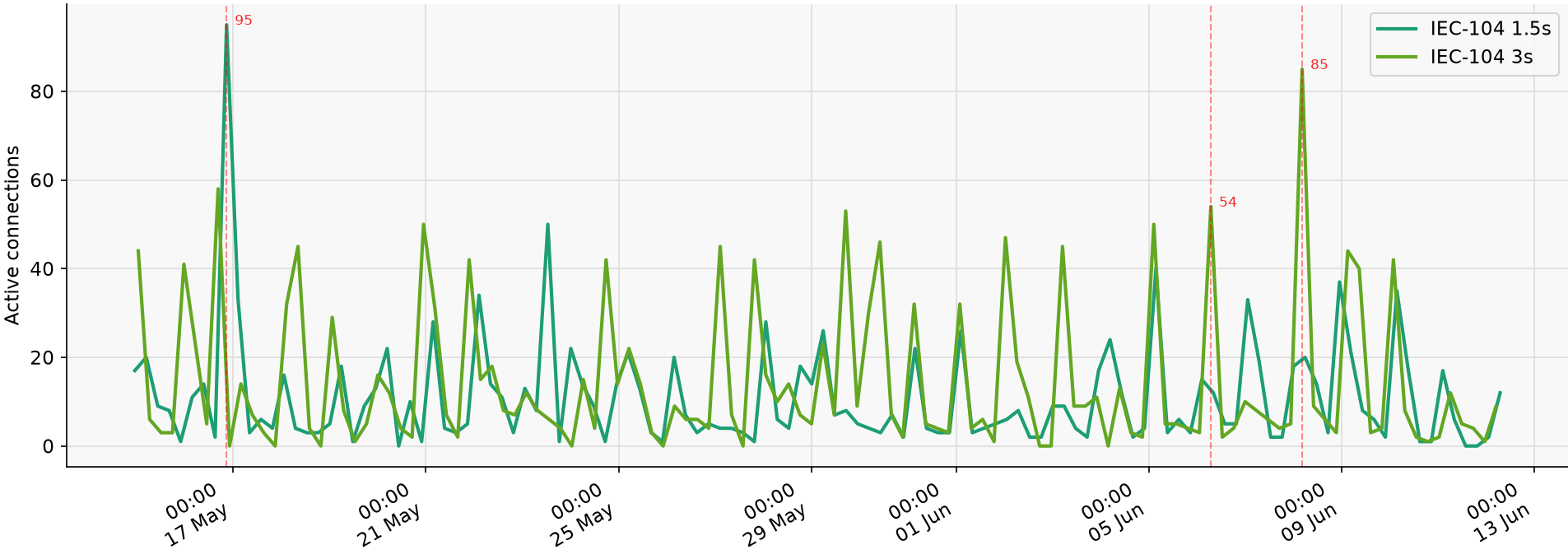}
\caption{IEC-104 active connections over time}
\label{fig:connections_overtime_iec104}
\end{figure}

\begin{figure}[!htb]
\centering
\includegraphics[width=0.75\linewidth]{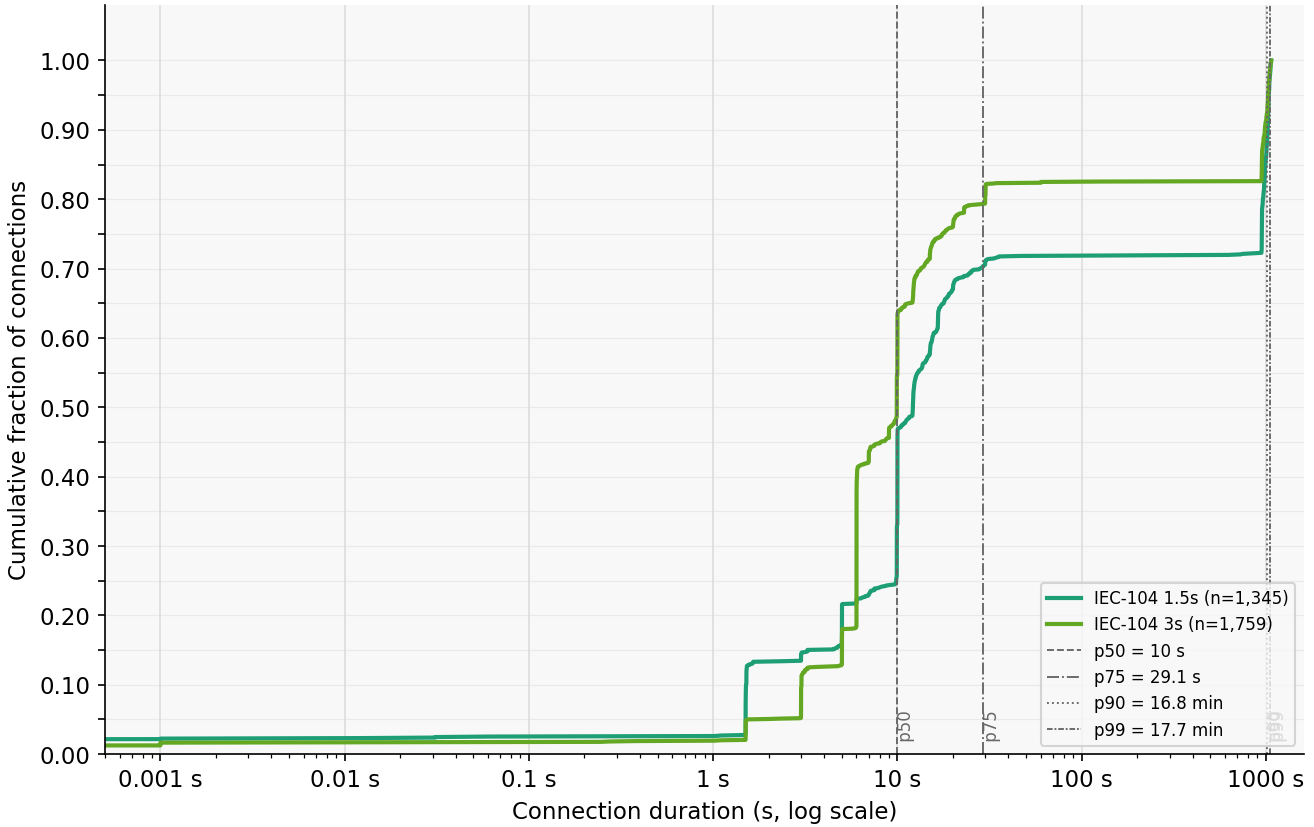}
\caption{IEC-104 connection-duration CDF}
\label{fig:cdf_overlap_iec104}
\end{figure}

Both IEC-104 tarpits experience temporal peaks in active connections at the same time, though less closely synchronized than the Modbus tarpits (Figure~\ref{fig:connections_overtime_iec104}). Their connection-duration distributions (Figure~\ref{fig:cdf_overlap_iec104}) show disconnect spikes around 1.5\,s and 5\,s for both tarpits, with Tarpit 4 additionally showing larger spikes at 3\,s and 6\,s. Tarpit 3 shows a further spike near 10\,s, and both tarpits show later spikes around 30\,s and 1000\,s.

IEC-104 shows a near-total absence of protocol-state-machine violations: no client was disconnected for an invalid sequence number or excessive request rate, and the only application-layer command observed on either tarpit was the type-100 General Interrogation. Full compliance with the expected specification sequence is nonetheless partial rather than universal: 57\% of Tarpit 3 (1.5s) sessions and 71\% of Tarpit 4 (3s) sessions adhere fully to it. A silent-versus-requesting split also holds for IEC-104, with silent connections persisting toward the 1000-second mark (Figure~\ref{fig:req_vs_silent_cdf_iec104}).

% Dropped from the active draft, kept for reference in case reinstated:
% \begin{figure}[!htb]
%     \centering
%     \includegraphics[width=\linewidth]{Pictures/33a_iec_timing_flags.png}
%     \caption{IEC-104 protocol-timing/behavior flags over the full deployment. Grey: fully specification-compliant; green: I-frame received before \texttt{STARTDT} acknowledgment; purple: no \texttt{STARTDT} and no activity; pink: no \texttt{STARTDT}, only \texttt{TESTFR} sent; teal: no client \texttt{TESTFR} despite the specification's maximum timeout~\cite{ds_en_60870_5_104}. ``Fast disconnect'' denotes connections under 1\,s.}
%     \label{fig:iec_timing}
% \end{figure}

\subsection{Cross-Protocol Comparison}
\label{subsec:cross-protocol}

Protocol choice compounds the pattern seen within each protocol. Over the overlap period, IEC-104 draws 26.5\% more average sessions per tarpit than Modbus (1212 vs.\ 958.3) yet accumulates 84.5\% less average stall time per tarpit (75.39 vs.\ 487.1 hours). Pairwise comparisons range from 298\% to 650\% more stall time for Modbus tarpits over IEC-104 tarpits.

This per-tarpit average, however, does not hold once volume is summed across each protocol's full complement of tarpits. Over the full deployment period, aggregate connection volume favors Modbus: 3{,}605 sessions (1{,}537+1{,}092+976) against IEC-104's 3{,}104 (1{,}345+1{,}759), 16.1\% more, and 1{,}096 cumulative unique IP addresses against IEC-104's 943, 16.2\% more. The reversal follows from tarpit count: Modbus's three deployments each draw somewhat less traffic than either IEC-104 tarpit individually, but their number pushes cumulative scanning volume for the protocol above IEC-104's two-tarpit total. Peak concurrency shows the same aggregate skew: the three Modbus tarpits average 24.7 concurrent connections (20, 28, 26) against an identical 12 for both IEC-104 tarpits, more than double.

IEC-104 records near-zero hard protocol-state-machine violations: no client is disconnected for an invalid sequence number or an excessive request rate (Subsection~\ref{subsec:iec104-results}). Full compliance with the expected specification sequence nonetheless holds for only 57\% of Tarpit 3 sessions and 71\% of Tarpit 4 sessions, leaving 43\% and 29\% of sessions respectively partially non-compliant without triggering a hard violation; no analogous metric exists for Modbus.

\begin{figure}[!htb]
\centering
\includegraphics[width=0.75\linewidth]{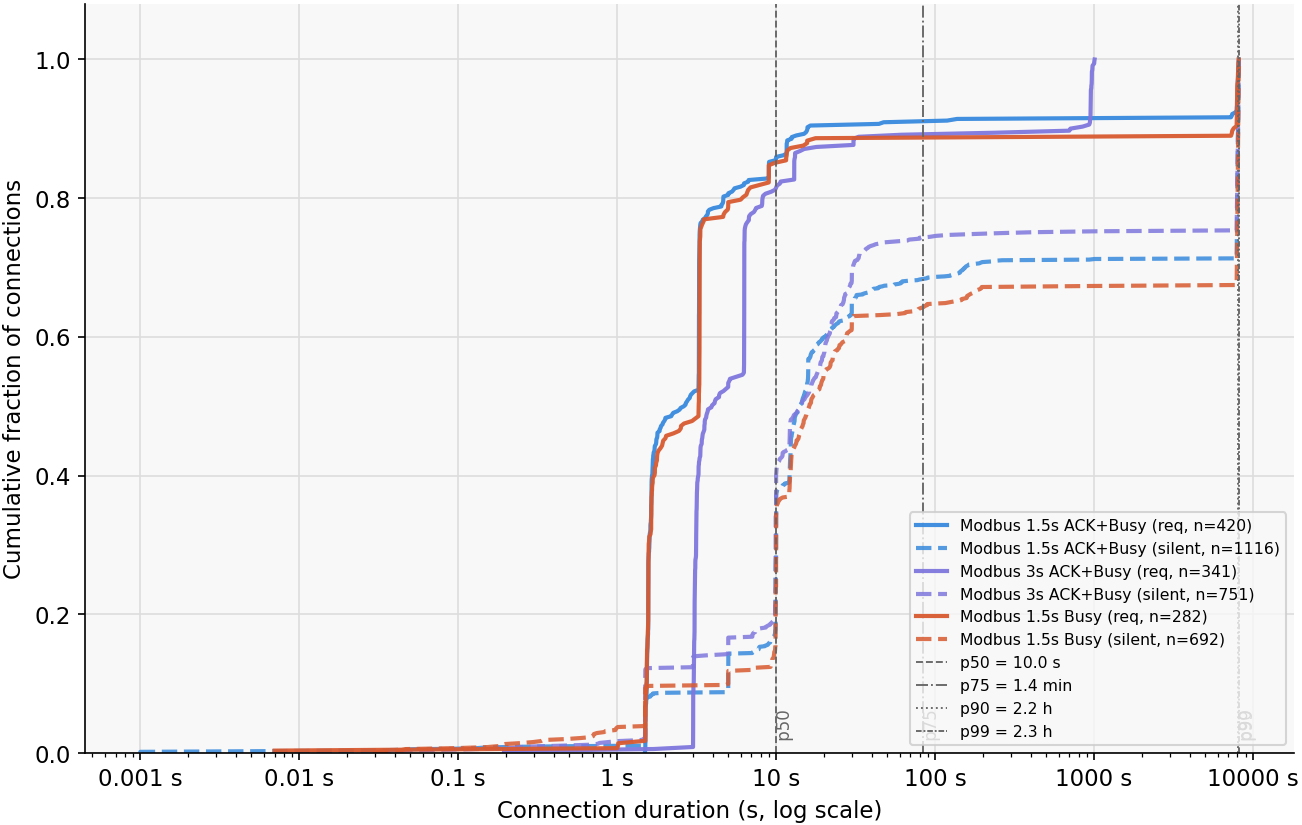}
\caption{Disconnect-time CDF, Modbus (requesting vs.\ silent)}
\label{fig:req_vs_silent_cdf_modbus}
\end{figure}

\begin{figure}[!htb]
\centering
\includegraphics[width=0.75\linewidth]{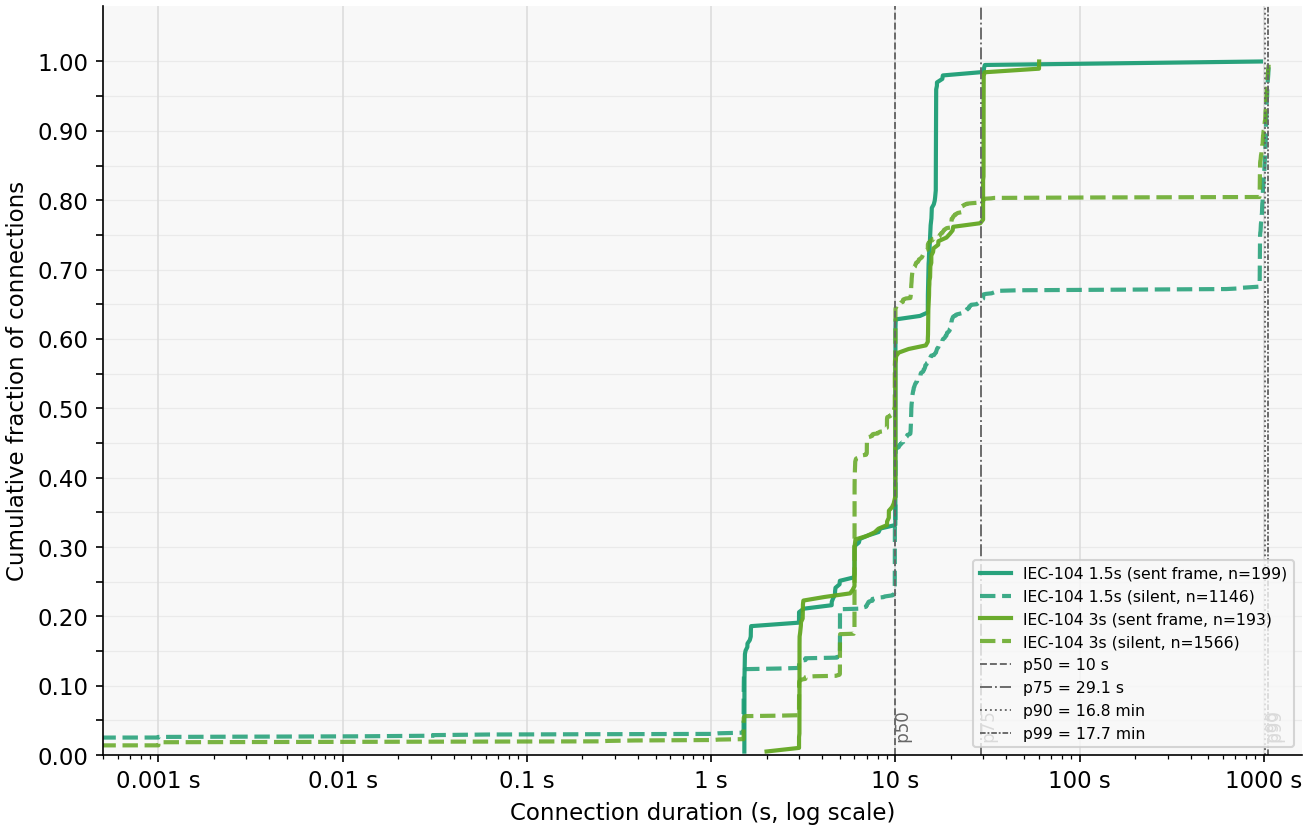}
\caption{Disconnect-time CDF, IEC-104 (requesting vs.\ silent)}
\label{fig:req_vs_silent_cdf_iec104}
\end{figure}

The two protocols nonetheless share several structural features. Both show the same non-monotonic relationship between induced delay and accumulated stall time: the shorter-delay tarpit accumulates more stall time than the longer-delay tarpit, for Modbus (Subsection~\ref{subsec:modbus-results}) and for IEC-104 (Subsection~\ref{subsec:iec104-results}) alike. Their connection-duration CDFs also share a timing signature, with a disconnect-spike cluster around 1.5\,s in both (Figure~\ref{fig:cdf_overlap} for Modbus, Figure~\ref{fig:cdf_overlap_iec104} for IEC-104), coinciding with each protocol's shorter-delay tarpit. A further parallel holds in client engagement: only 27.3\%--31.2\% of Modbus connections ever issue a valid request (Subsection~\ref{subsec:modbus-results}), and a comparable silent-versus-requesting split holds for IEC-104 sessions, with requesting and silent populations showing distinct disconnect-time distributions in both protocols (Figures~\ref{fig:req_vs_silent_cdf_modbus}, \ref{fig:req_vs_silent_cdf_iec104}).

Reconnection behavior is the one dimension where IEC-104 draws more repeat engagement than Modbus. The Modbus tarpits show reconnection rates of 32.9\% (Tarpit 1 and 5) and 34.1\% (Tarpit 2), while the IEC-104 tarpits show 48.2\% (Tarpit 3) and 48.9\% (Tarpit 4), 14--16 percentage points higher.

\subsection{Threat-Intelligence Correlation}
\label{subsec:threat-intel}

This subsection examines how GreyNoise’s malicious classification correlates with stall time, protocol compliance, and client persistence across both protocols.

\begin{table}[ht]
    \centering
    \caption{Session-level summary, by GreyNoise classification of the connecting IPs.}
    \label{tab:gn-class}
    \vspace{2pt}
    \footnotesize
    \setlength{\tabcolsep}{6pt}
    \resizebox{\linewidth}{!}{%
    \begin{tabular}{lc cc cc cc}
        \toprule
        & & \multicolumn{2}{c}{\textbf{Malicious}} & \multicolumn{2}{c}{\textbf{Benign}} & \multicolumn{2}{c}{\textbf{Unknown}} \\
        \cmidrule(lr){3-4}\cmidrule(lr){5-6}\cmidrule(lr){7-8}
        \textbf{Protocol} & \textbf{ID} & \textbf{tw} & \textbf{sn} & \textbf{tw} & \textbf{sn} & \textbf{tw} & \textbf{sn} \\
        \midrule
        \multirow{3}{*}{Modbus} & 1 & 720.0 (89.9\%) & 977 (63.6\%) & 80.2 (10.0\%) & 434 (28.3\%) & 0.3 (0.0\%) & 125 (8.1\%) \\
         & 2 & 416.5 (97.4\%) & 636 (58.2\%) & 8.8 (2.1\%) & 337 (30.9\%) & 2.5 (0.6\%) & 119 (10.9\%) \\
         & 5 & 504.6 (87.3\%) & 576 (59.1\%) & 70.7 (12.2\%) & 279 (28.6\%) & 2.5 (0.4\%) & 120 (12.3\%) \\
        \midrule
        \multirow{2}{*}{IEC-104} & 3 & 95.4 (88.8\%) & 811 (60.3\%) & 11.6 (10.8\%) & 384 (28.5\%) & 0.4 (0.4\%) & 150 (11.2\%) \\
         & 4 & 78.9 (88.8\%) & 876 (49.8\%) & 9.5 (10.7\%) & 682 (38.8\%) & 0.5 (0.6\%) & 201 (11.4\%) \\
        \bottomrule
    \end{tabular}%
    }
    \par\vspace{3pt}
    \noindent{\textbf{tw}: stall time. \quad \textbf{sn}: session/disconnect count.}
\end{table}

Table~\ref{tab:gn-class}'s session-share column ranks the tarpits by relative malicious exposure: Tarpit 1 highest at 63.6\%, then Tarpit 3 (60.3\%), Tarpit 5 (59.1\%), Tarpit 2 (58.2\%), and Tarpit 4 lowest at 49.8\%, a 13.8-percentage-point spread end to end. Pooled by protocol, malicious-tagged IPs account for 60.8\% of Modbus sessions against 54.3\% of IEC-104 sessions.

Every tarpit's malicious time-share also exceeds its malicious session-share, and the disproportion is sharpest exactly where the malicious minority is smallest: from 1.41 times (Tarpit 1, 89.9\%/63.6\%) up to 1.78 times (Tarpit 4, 88.8\%/49.8\%), falling monotonically as session-share rises across the five tarpits.

\begin{figure}[!htb]
\centering
\includegraphics[width=0.75\linewidth]{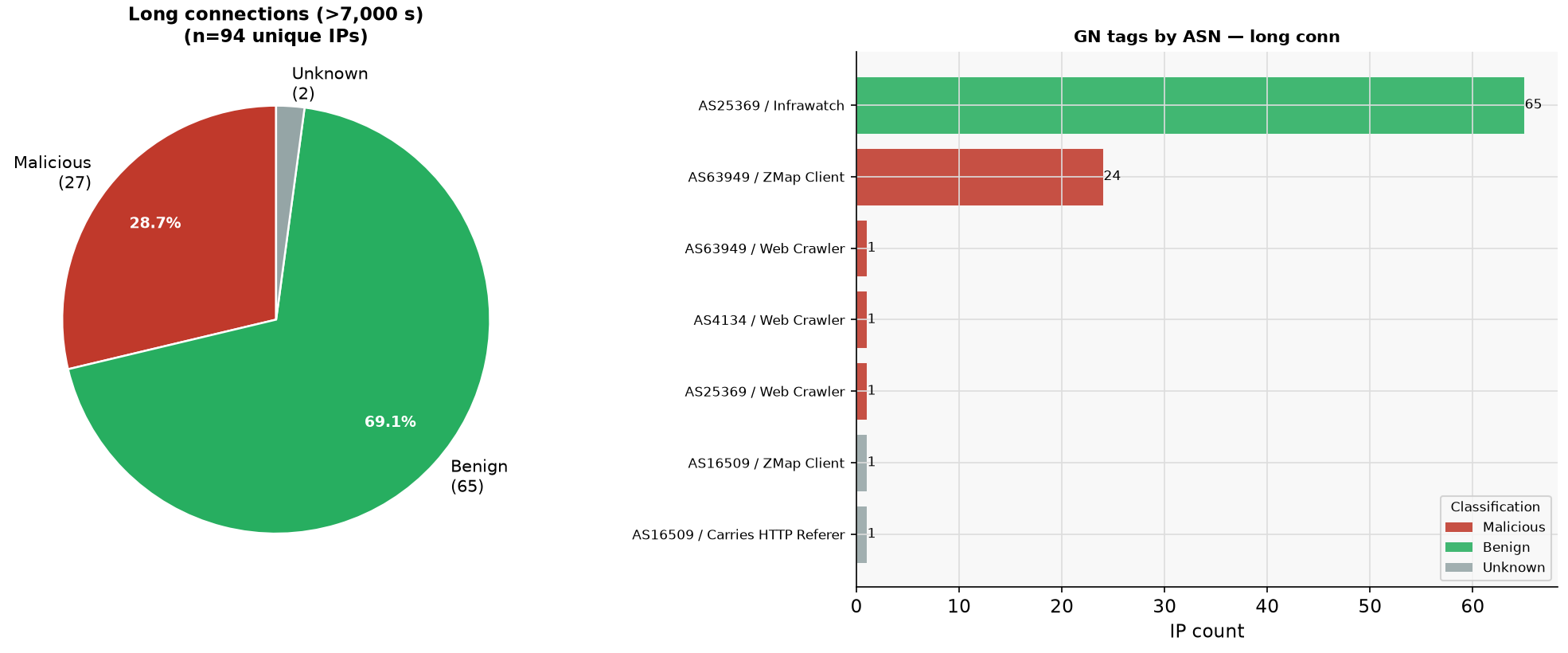}
\caption{GreyNoise classification of long-duration Modbus connections}
\label{long_conn_modbus}
\end{figure}

Excluding known scanning services (Shodan.io, ShadowServer.org, Infrawatch), every Modbus IP issuing more than three requests across sessions and every IP sustaining long-duration connections is tagged malicious (Figure~\ref{long_conn_modbus}), consistent with every one of the twenty most active clients across all five tarpits, spanning both protocols, carrying a malicious tag.

A function-code breakdown of the same Modbus traffic sharpens this picture. Every one of the 84 FC03 (Read Holding Registers) requests recorded across the three tarpits (Table~\ref{tab:modbus-fc-all}) originated from a malicious-tagged IP, though from only a handful of distinct clients; FC17 (Report Server ID), the second-largest category at 304 requests, rarely came from malicious-tagged IPs; and FC43 (Read Device Identification), the dominant category at 1{,}031 requests, was used broadly across all three GreyNoise classifications rather than concentrating in any one.

Session duration diverges just as sharply by classification. Over the overlap period, more than 80\% of benign-tagged Modbus sessions disconnect within 10 seconds, against only 20\% of malicious-tagged sessions in the same window. The remaining malicious sessions spread across a rounded disconnect curve out to 1{,}000 seconds, and roughly a third persist to the 2.2--2.3-hour mark before disconnecting, consistent with the concentration of long-duration malicious connections beyond the roughly 7{,}000-second mark noted above.

Among IEC-104 sessions that do not fully complete the expected specification sequence, the malicious-classification rate increases by 24.7\% for Tarpit 3 and by 29.4\% for Tarpit 4 relative to compliant sessions.

% Dropped from the active draft, kept for reference in case reinstated:
% \begin{figure}[ht]
% \centering
% \begin{subfigure}{0.48\textwidth}
%     \centering
%     \includegraphics[width=\linewidth]{Pictures/33d_iec_gn_class_tp3_compliant.png}
%     \caption{Compliant connections}
% \end{subfigure}
% \hfill
% \begin{subfigure}{0.48\textwidth}
%     \centering
%     \includegraphics[width=\linewidth]{Pictures/33d_iec_gn_class_tp3_broke.png}
%     \caption{Non-compliant connections}
% \end{subfigure}
% \caption{GreyNoise classification of connecting IPs, split by IEC-104 protocol compliance, Tarpit 3 (1.5\,s).}
% \label{fig:33d-tp3}
% \end{figure}
%
% \begin{figure}[ht]
% \centering
% \begin{subfigure}{0.48\textwidth}
%     \centering
%     \includegraphics[width=\linewidth]{Pictures/33d_iec_gn_class_tp4_compliant.png}
%     \caption{Compliant connections}
% \end{subfigure}
% \hfill
% \begin{subfigure}{0.48\textwidth}
%     \centering
%     \includegraphics[width=\linewidth]{Pictures/33d_iec_gn_class_tp4_broke.png}
%     \caption{Non-compliant connections}
% \end{subfigure}
% \caption{GreyNoise classification of connecting IPs, split by IEC-104 protocol compliance, Tarpit 4 (3\,s).}
% \label{fig:33d-tp4}
% \end{figure}

% Dropped from the active draft, kept for reference in case reinstated:
% \begin{figure}[!htb]
%     \centering
%     \includegraphics[width=\linewidth]{Pictures/31c_gn_classification_high_req.png}
%     \caption{GreyNoise classification (by ASN) of unique Modbus IPs issuing more than three requests across sessions, excluding known scanning services.}
%     \label{high_req_modbus}
% \end{figure}

\begin{figure}[!htb]
\centering
\includegraphics[width=0.75\linewidth]{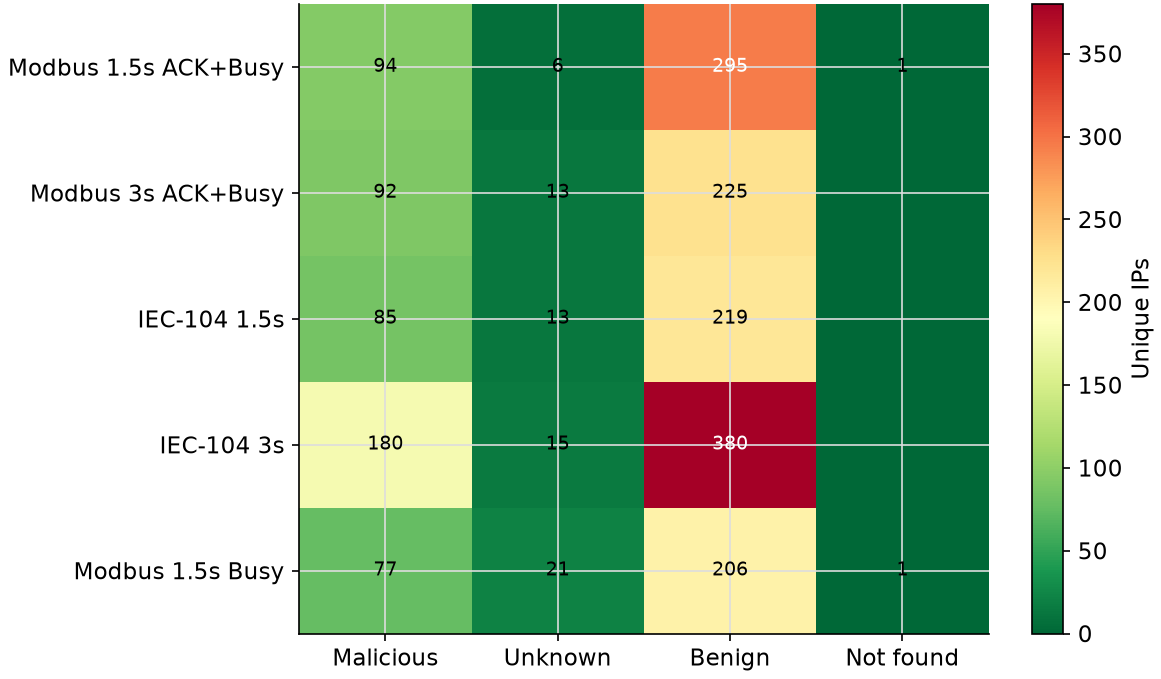}
\caption{Unique-IP counts by tarpit and GreyNoise classification}
\label{fig:gn_cross_tarpit_heatmap}
\end{figure}

% Dropped from the active draft, kept for reference in case reinstated:
% \begin{figure}[!htb]
%     \centering
%     \includegraphics[width=\linewidth]{Pictures/10b_ip_heatmap_by_classification.png}
%     \caption{Connection counts for the top 20 most active clients across all five tarpits, colored by GreyNoise classification.}
%     \label{fig:ip_heatmap_gn}
% \end{figure}

\section{Discussion}
\label{sec:discussion}

\subsection{Protocol Compliance and Threat-Intelligence Corroboration}
\label{subsec:disc-drift}

We treat GreyNoise tags as a corroboration heuristic, not ground truth, reflecting scanner reputation from independent telemetry rather than a verified label per session. Read cautiously, Section~\ref{subsec:threat-intel} (Table~\ref{tab:gn-class}) shows a real disparity: malicious-tagged IPs are only 22--30\% of unique connecting addresses across the five tarpits, yet account for 87.3--97.4\% of total stall time. Their session share is already a majority or plurality on its own (49.8--63.6\% per tarpit; 60.8\% pooled for Modbus, 54.3\% for IEC-104). It is the address-versus-stall-time gap, not the session share, that constitutes the minority-owns-majority pattern here. One plausible reading: a small set of persistent, reputation-flagged sources drives most engagement -- though tagging itself may simply correlate with the persistence that produces long stalls.

A second signal concerns compliance drift. IEC-104 shows near-zero hard state-machine violations alongside only 57\%/71\% full-sequence compliance for Tarpits 3 and 4 (\ref{subsec:iec104-results}). Non-compliance here means soft drift, such as I-frames preceding STARTDT acknowledgment, not malformed traffic. Among non-fully-compliant sessions, the malicious-classification rate is 24.7\% and 29.4\% higher, respectively, than among compliant sessions (\ref{subsec:threat-intel}). This may reflect deliberate fingerprinting -- probing state-machine edges before committing further effort -- or naive automation that never fully implements the handshake. The data cannot disentangle these. Hard-violation detection alone would catch almost none of this: soft compliance drift is the more sensitive discriminator here, and could feed OT intrusion-detection heuristics as a candidate feature, though legitimate non-standard implementations could produce identical drift benignly.

Modbus shows an analogous but distinct signal: not sequencing, but which function code a client chooses (\ref{subsec:threat-intel}, Table~\ref{tab:modbus-fc-all}). Function Code 43 (Read Device Identification), the dominant request at roughly 73\% of Modbus traffic, appears broadly across malicious, benign, and unknown sources: a generic probe any scanner might send, malicious or not. Function Code 17 (Report Server ID, 304 requests) was likewise rarely malicious-tagged. Function Code 03 (Read Holding Registers), a rarer sensor-level read, appeared only 84 times, yet every instance was malicious-tagged, from just a handful of distinct clients -- too small a sample to generalize with confidence. As with IEC-104, anomalies concentrate on the suspicious side while dominant traffic stays tag-agnostic. Here the signal is command choice, not sequence compliance, and FC03 remains fully valid, not malformed.

\subsection{Delay and Stall-Time Sensitivity}
\label{subsec:disc-retention}

The delay-to-stall-time association is non-monotonic, and the same association was observed across two structurally different tarpitting strategies. Modbus's shorter-delay tarpits (1.5s) accumulate 75.7--80.3\% more total stall time than its longer-delay tarpit (3s); IEC-104's shorter-delay tarpit (1.5s) accumulates 20.7\% more than its longer-delay tarpit (3s) (\ref{subsec:modbus-results}, \ref{subsec:iec104-results}). A second metric points the same direction in both protocols: mean connection duration is also longer on the shorter-delay tarpit, by 7.77 and 12.08 minutes over the full deployment period for Modbus Tarpits 1 and 5 versus Tarpit 2, and by 59.7\% for IEC-104 Tarpit 3 versus Tarpit 4. Two different tarpitting mechanisms -- Modbus's Exception Code strategy and IEC-104's state-machine stalling -- and two independent metrics pointing the same direction is what makes this an association worth reporting, not an artifact of one implementation. It stops short of an isolated causal effect of delay, however: each tarpit occupies its own public IP, delay was not rotated or replicated across IPs (\ref{sec:eval-tarpits}), and IP-specific scanning patterns (e.g., which scanning services happen to target a given address) could contribute to the gap alongside delay itself. The Modbus comparison also rests on heavily right-skewed data -- mean connection duration (1,877--2,136\,s) far exceeds the median (10\,s) for all three tarpits (\ref{subsec:modbus-results}) -- so a relatively small number of long sessions likely drive most of the measured difference in total stall time.

This effect concentrates in the minority of traffic that engages at the protocol level: only around 27--31\% of Modbus connections ever issue a request (\ref{subsec:modbus-results}), and raw connection counts otherwise include a large non-requesting population that delay tuning does not touch. A similar split is known for IEC-104 (\ref{subsec:iec104-results}). Delay reshapes already-engaged clients' behavior, not the full incoming volume.

\subsection{Reconnection Rate and Traffic Analysis}
\label{subsec:disc-reconnection}

Reconnection rate is a distinct engagement dimension from stall time: a protocol can be revisited often without being expensive per visit, or vice versa. IEC-104 tarpits show markedly higher reconnection rates (48.2--48.9\%) than Modbus tarpits (32.9--34.1\%, \ref{subsec:cross-protocol}). Two further patterns bear on how that traffic should be read. Connection activity across the deployment period follows a cyclical, roughly daily peak-and-trough shape (\ref{fig:connections_overtime_modbus}, \ref{fig:connections_overtime_iec104}) rather than a flat or purely random arrival process, consistent with scheduled, looped scanning infrastructure of the kind large-scale internet-scanning services routinely run. The longest engaged (non-silent) Modbus sessions, separately, cluster near a duration that lines up with standard TCP keep-alive default timing -- roughly two hours plus periodic probe overhead (\ref{fig:cdf_overlap}) -- rather than any duration the tarpit itself induces, suggestive of default OS- or tool-level keep-alive behavior sustaining the session rather than deliberate human persistence. Together, reconnection rate, daily periodicity, and keep-alive-aligned session ceilings point toward scheduled/automated tooling as the dominant driver of sustained engagement. This reading must be hedged: none of these signals individually rules out a human-in-the-loop process, such as an analyst periodically re-triggering the same tool or manually following up on a scanning service's output.

\subsection{Per-Protocol Engagement and Aggregated Stall Time}
\label{subsec:disc-fleet}

Per-tarpit averages and aggregate totals answer different questions. Section~\ref{subsec:cross-protocol}'s data show them diverging here: per tarpit, IEC-104 draws 26.5\% more average sessions than Modbus (1212 vs.\ 958.3), while Modbus accumulates roughly 546\% more average stall time per tarpit than IEC-104 (487.1 vs.\ 75.39 hours; pairwise, 298--650\% more for individual Modbus tarpits over individual IEC-104 tarpits). Summed across each protocol's full deployment, this pattern reverses: three Modbus tarpits draw 3,605 total sessions and 1,096 cumulative unique IPs against two IEC-104 tarpits' 3,104 sessions and 943 IPs, 16.1\% and 16.2\% more. The reversal traces to tarpit count, not any single Modbus tarpit outperforming an IEC-104 tarpit: each Modbus tarpit alone draws less traffic than either IEC-104 tarpit, but three combined exceed two. Peak concurrency shows the same skew -- Modbus's three tarpits average 24.7 concurrent connections against an identical 12 for both IEC-104 tarpits, though with only two data points the equality may be coincidental, not a structural ceiling.

Per-node efficiency and total defensive yield are separate metrics that can favor different protocols -- a distinction any deception-deployment composition decision should track explicitly. With only three and two tarpits, we cannot say whether this reversal holds at larger deployments or different protocol mixes, nor why IEC-104's per-tarpit session and stall-time trends run opposite directions.

\section{Limitations and Mitigations}
\label{sec:limitations}

Tarpitting is structurally fingerprintable. It requires a specific, repeatable sequence of events to function, and a sufficiently capable scanner can detect that sequence; this is a property of the technique, not a defect of this deployment. Mladenov et al.~\cite{glittersGold} found industrial deception technologies detected at rates reaching 92\% in some protocol categories, and Yaben et al.~\cite{yaben5974783measuring} developed techniques specifically to fingerprint noise and uncommon connections behaviors, in OT environments. A second limitation compounds it: all five tarpits ran on same-region instances, each on its own public IP, with no configuration rotation across IPs (Section~\ref{sec:eval-setup}). Per-IP conditions are not guaranteed homogeneous. Two same-protocol tarpits differing only in configured delay showed a $\approx$6\% discrepancy in daily unique-IP counts, consistent with IP-specific exposure variance rather than a controlled comparison.

This hosting choice was deliberate: separate IPs, rather than a shared host, limit an adversary's ability to correlate the tarpits and fingerprint the deployment as a whole. Varying the configured reply delay (1.5s vs.\ 3s) produced no observed difference in threat-intelligence classification: no tarpit was flagged as suspicious, or identified as a tarpit, at a rate that tracked its delay setting. The narrow 22--30\% malicious-classification band observed for connecting \emph{client} IPs across all five tarpits (\ref{subsec:disc-drift}) points the same way, though it measures a different signal: client reputation, not the tarpits' own.
\section{Conclusion and Future Work}
\label{sec:conclusion}

We designed and deployed application-layer tarpits for Modbus and IEC-104, using protocol-native semantics: Modbus Exception Codes \texttt{0x05}/\texttt{0x06} and prolonged residence in IEC-104's connected state machine. To our knowledge, these are the first protocol-native application-layer tarpits for OT/ICS evaluated in a live Internet deployment, extending application-layer tarpitting beyond the generic IoT services targeted by prior work. Five variants ran for 24 days on the open internet, three Modbus and two IEC-104, logging 6,709 sessions and 2,039 cumulative per-tarpit unique-IP observations over 2,000 hours of engagement. GreyNoise enrichment attributes 87--97\% of that stall time to independently classified malicious sources, though only 22--30\% of connecting addresses carry that tag: a small, persistent population drives most engagement.

Two findings complicate a simple engagement-maximizing view of deception. Rare Modbus function-code requests and soft IEC-104 compliance drift both correlate with malicious classification, a fingerprinting signal beyond dwell time. The delay-engagement relationship is also non-monotonic: shorter delays yielded more stall time and longer connections than longer ones in both protocols. With only five variants deployed, per-protocol efficiency trends remain unconfirmed. 

Two directions follow from these findings. First, our deployment covers only Modbus and IEC-104; OT/ICS spans a wider set of legacy protocols built with little security in mind, and extending the same tarpitting approach across that protocol pool is a direct, low-risk expansion of this work. Second, and more pressing for defenders: Section~\ref{sec:limitations} shows tarpitting is structurally fingerprintable, and prior work has already demonstrated detection of industrial deception technologies~\cite{glittersGold,yaben5974783measuring}. Hardening the deception layer against exactly this adversary is the natural next step before wider OT tarpit deployment.

\clearpage

\clearpage
% optional clearing of the page
%\cleardoublepage
%\newpage
\appendix
%\section*{Ethical Considerations}
% \input{appendices/ethics.tex}
% \input{appendices/reproducibility}
\section{Open Science}
\label{sec:open-science}

For the sake of reproducibility and the artifacts check, we provide the tarpit code and the material used from the tarpit logs: a Modbus TCP tarpit, an IEC~60870-5-104 (IEC-104) tarpit, and the raw connection logs from all five deployments described in the evaluation. These logs allow independent verification of the reported measurements. Source IP addresses are irreversibly pseudonymized through a salted hash, applied consistently across all five logs so repeated connections from the same source remain linkable across deployments. This supports the reconnection-rate analysis in Section~\ref{sec:discussion} without exposing real IP addresses. Timestamps and protocol-level fields, including function codes, frame types, information object addresses, and exception codes, are left unmodified. Code and logs are available at \url{https://anonymous.4open.science/r/StallGrid-A765/}.
\section{Ethical Considerations}
\label{app:ethics}

The EU NIS 2 Directive mandates proactive defense measures from covered entities to prevent and minimize incident impact, motivating tarpitting as a compliant strategy \cite{nis2_directive_2022}. Both tarpits comply with applicable unauthorized-access and IT-disruption statutes by construction, never disrupting third-party operations. Connecting client IP addresses constitute personal data under Article 4 of the GDPR \cite{gdpr_regulation_2016}, and Article 89 exempts such processing for scientific research \cite{gdpr_article_89}. The collected data is processed solely for this research, by its authors.

The ethical justification follows a self-defense and proportionality framework from prior work on defensive cyber deception \cite{ExploringEthics}: harm to a connecting scanner (wasted time and resources probing a non-functional endpoint) does not exceed the harm prevented (reconnaissance against real critical infrastructure). The goal is characterizing aggregate active engagement behavior and assessing tarpit defensive potential, not attributing individual attacks or deanonymizing operators. By design (Section~\ref{sec:design}), both tarpits accept only client-initiated connections and never originate outbound traffic. Logged data is restricted to connection metadata: timestamps, client IP addresses, and protocol-level fields (function codes, frame types, information object addresses, exception codes). No request-supplied payload is executed or relayed: the tarpits implement only enough protocol state-machine logic to sustain a connection, without simulating device functionality or control.

Regarding network bandwidth impact, prior application-layer tarpit deployments report negligible aggregate bandwidth, well under a single HD video stream, since tarpit replies are short, protocol-compliant messages rather than retransmitted payloads \cite{safargalieva2026eventhorizon}. OT networks are, by definition, bandwidth-constrained industrial environments, so this deployment is not an exception.
\bibliographystyle{splncs04}
\bibliography{references}

\end{document}